\documentstyle[pra,aps]{revtex}
\begin{document}
\draft
\title{Nonclassical effects in a driven atoms/cavity system in the presence of 
arbitrary driving field and dephasing}
\author{J. P. Clemens and P. R. Rice}
\address{Department of Physics, Miami University, Oxford, Ohio 45056}
\date{\today}
\maketitle
\begin{abstract}
We investigate the photon statistics of light transmitted from a driven 
optical cavity containing one or two atoms interacting with a single mode of 
the cavity field.  We treat arbitrary driving fields with emphasis on 
departure from previous weak field results.  In addition effects of dephasing 
due to atomic transit through the cavity mode are included using two different 
models.  We find that both models show the nonclassical correlations are quite 
sensitive to dephasing.  The effect of multiple atoms on the system dynamics 
is investigated by placing two atoms in the cavity mode at different 
positions, therefore having different coupling strengths.
\end{abstract}

\section{Introduction}
In this paper we report on extensions to previous work on dynamical cavity QED 
effects in the photon statistics of transmitted light from a driven optical 
cavity coupled to an ensemble of two level atoms.  Much work has been done on 
structural cavity QED effects such as energy level shifts and the modification 
of spontaneous emission rates.  These structural effects can be seen to arise 
from semiclassical models.  In addition work has been done on dynamical 
effects where the coupling between the cavity field and atoms has a 
significant effect on the evolution of the system, in particular in the strong 
coupling regime where a single quantum of energy and hence single quantum 
fluctuations give rise to nontrivial dynamics.  In this regime the field 
can neither be viewed as mildly perturbed by the atoms (good cavity limit), nor are 
the atoms mildly perturbed by the field (bad cavity limit).  For a review of 
the work on structural and dynamical effects in cavity QED, see Ref. 
\cite{Berman}.

The problem of a single two level atom coupled to a single mode field was 
originally studied by Jaynes and Cummings \cite{jaynes&cummings} and extended 
to many atoms by Tavis and Cummings \cite{tavis&cummings1,tavis&cummings2}.  
These models have been extended in recent theoretical work to include 
spontaneous emission and cavity field decay  \cite{IEEE,optcomm}, and atomic 
transit time broadening and detunings \cite{fatpaper}.  Nonclassical 
correlations in photon statistics which violate a Schwarz inequality have been 
predicted for this system, including photon antibunching [defined here as 
$g^{(2)}(0)_+ > g^{(2)}(0)$], and sub-Poissonian statistics 
[$g^{(2)}(0) < 1$].  Also other effects have been predicted which we refer to 
as overshoots and undershoots [$|g^{(2)}(\tau) - 1| > |g^{(2)}(0) - 1|$ where 
$g^{(2)}(\tau)$ is the normalized second order correlation function].  
Examples of these nonclassical correlations from previous weak field results 
are shown in Fig. \ref{fig:weak}.  Figure \ref{fig:weak}(a) shows photon 
antibunching and sub-Poissonian statistics, (b) shows an overshoot violation, 
and (c) shows an undershoot violation.

Photon antibunching has been seen experimentally in this system by Rempe et. 
al. \cite{Rempe}.  Overshoot violations have recently been seen by Mielke, 
Foster, and Orozco \cite{Luis}.  In general the theory matches the experiments 
in terms of qualitative behavior while the quantitative size of the 
nonclassical effect does not.  This has led us to consider complications in 
the experiments which may be responsible for the discrepancy including 
deviations from the weak field limit, and dephasing due to the atoms entering 
and leaving the cavity.  The experiments use an atomic beam to introduce atoms 
into the cavity so that the time of flight across the mode is on the order of 
ten spontaneous emission lifetimes \cite{Luis}.  We expect that dephasing due 
to atomic traversal of the cavity will have a detrimental effect on 
nonclassical correlations.  In addition deviations from the weak driving field 
limit and interactions with 'spectator' atoms far from the mode waist may be 
important.  These effects are investigated in this paper by numerically 
solving the master equation for the system and by quantum trajectory 
simulations.  Rather than investigating all possible effects at once, we 
isolate them and try to understand what is most critical.

The general outline of this paper is as follows.  In Section \ref{sec:model} 
we present the physical model of the system under investigation and describe 
the methods of solution.  In Section \ref{sec:strong} we discuss the photon 
statistics of the transmitted light outside the weak field limit.  Section 
\ref{sec:dephasing} presents two models of atomic transit dephasing and the 
resulting photon statistics.  In Section \ref{sec:two atoms} we include 
effects of a spectator atom with a coupling which is a fraction of the 
maximum coupling, and finally we conclude in Section \ref{sec:conclude}.

\section{Physical model}
\label{sec:model}
The system under investigation is an extension of the Jaynes-Cummings 
Hamiltonian which includes effects of atomic and cavity field decay as well as a 
coherent driving field.  A schematic diagram of the system is shown in Fig. 
\ref{fig:diagram}.  The field and atomic Hamiltonians are given by
\begin{equation}
H_F = \hbar\omega_c a^\dagger a
\end{equation}
\begin{equation}
H_A = \sum_j \hbar\omega_a\sigma_z^j
\end{equation}
and the atom-field interaction in the rotating wave approximation is given by
\begin{equation}
H_{AF} = \sum_j i\hbar g_j \left ( a^\dagger \sigma_-^j - a \sigma_+^j \right).
\end{equation}
The cavity field creation and annihilation operators are $a^\dagger $ and $a$ 
respectively and $\sigma_\pm^j$ and $\sigma_z^j$ are Pauli operators for the 
$j$th two-level atom.  The atom-field coupling strength is determined by
\begin{equation}
g_j = \mu \left ( \frac{\omega_c}{2\hbar\epsilon_0 V} \right )^{1/2} \sin 
kz_j
\end{equation}
where $\mu$ is the dipole transition matrix element between the two atomic 
states, $V$ is the cavity mode volume, and $\sin kz_j$ takes into account the 
position of the atom in the mode.  In previous work it was assumed that the 
atoms sit at antinodes of the field where the coupling is a maximum.  Here we 
allow the atoms to be placed anywhere in the mode so that a range of 
couplings are allowed for different atoms.  The cavity field is driven by a 
classical laser field with the driving field-cavity field coupling described 
by the Hamiltonian 
\begin{equation}
H_L = i\hbar E \left ( a^\dagger e^{-i\omega_d t} - a e^{i\omega_d t} \right )
\end{equation}
where $E$ is the classical laser intensity scaled such that $E/\kappa$ is the 
photon flux injected into the cavity.  Throughout we assume the 
atom, cavity field, and driving field are on resonance ($\omega_0 \equiv 
\omega_c = \omega_a = \omega_d$).

Dissipation in the system gives rise to nontrivial irreversible dynamics.  
Cavity field damping and atomic population and polarization decay are 
described by superoperators acting on the density matrix of the system which 
are derived using standard methods \cite{gardiner,Howardbig}.  Cavity field 
damping is described by
\begin{equation}
{\cal L}_F \rho = \kappa \left ( 2a\rho a^\dagger - a^\dagger a\rho - \rho
a^\dagger a \right )
\end{equation}
where $\kappa$ is the rate of cavity field damping.  Atomic population and 
polarization decay are described by
\begin{equation}
{\cal L}_A \rho = \frac{\gamma}{2} \sum_j \left (2\sigma_-^j\rho\sigma_+^j -
\sigma_+^j\sigma_-^j\rho - \rho\sigma_+^j\sigma_-^j \right )
\end{equation}
where $\gamma$ is the spontaneous emission rate of an atom.  The full master 
equation in the Born-Markov approximation is then
\begin{equation}
\dot \rho = -\frac{i}{\hbar}\left [ H_A + H_F + H_{AF} + H_L, \rho \right ] +
{\cal L}_F \rho + {\cal L}_A \rho \to {\cal L} \rho .
\end{equation}
A numerical solution of the master equation is carried out in the Fock state 
basis, and also a quantum trajectory simulation is developed from the master 
equation.

\subsection{Numerical solution of the master equation}
The master equation in the Fock state basis is
\begin{mathletters}
\begin{eqnarray}
\dot\rho_{n,+;m,+} & = & -g\sqrt{n+1}\rho_{n+1,-;m,+} -
g\sqrt{m+1}\rho_{n,+;m+1,-}
+ E\sqrt{n}\rho_{n-1,+;m,+} \\
&& + E\sqrt{m}\rho_{n,+;m-1,+} - E\sqrt{n+1}\rho_{n+1,+;m,+} 
- E\sqrt{m+1}\rho_{n,+;m+1,+} \nonumber \\
&&+ 2\kappa\sqrt{(n+1)(m+1)}\rho_{n+1,+;m+1,+} 
- \left [\kappa(n+m)+\gamma\right ]\rho_{n,+;m,+} \nonumber
\end{eqnarray}
\begin{eqnarray}
\dot\rho_{n,-;m,-} & = & g\sqrt{n}\rho_{n-1,+;m,-} + g\sqrt{m}\rho_{n,-;m-1,+}
+ E\sqrt{n}\rho_{n-1,-;m,-} \\
&& + E\sqrt{m}\rho_{n,-;m-1,-} - E\sqrt{n+1}\rho_{n+1,-;m,-} 
- E\sqrt{m+1}\rho_{n,-;m+1,-} \nonumber \\
&& + 2\kappa\sqrt{(n+1)(m+1)}\rho_{n+1,-;m+1,-} 
- \kappa (n+m)\rho_{n,-;m,-} + \gamma\rho_{n,+;m,+} \nonumber
\end{eqnarray}
\begin{eqnarray}
\dot\rho_{n,+;m,-} & = & -g\sqrt{n+1}\rho_{n+1,+;m,-} +
g\sqrt{m}\rho_{n,+;m-1,+} + E\sqrt{n}\rho_{n-1,+;m,-} \\
&& + E\sqrt{m}\rho_{n,+;m-1,-} - E\sqrt{n+1}\rho_{n+1,+;m,-} 
- E\sqrt{m+1}\rho_{n,+;m+1,-} \nonumber \\
&& + 2\kappa\sqrt{(n+1)(m+1)}\rho_{n+1,+;m+1,-} 
- \left [\kappa(n+m) - \gamma/2\right ]\rho_{n,+;m,-} \nonumber
\end{eqnarray}
\begin{equation}
\dot\rho_{n,-;m,+} = \dot\rho_{m,+;n,-}^*
\end{equation}
\end{mathletters}
where $\rho_{n,\pm;m,\pm} = \langle n,\pm|\rho |m,\pm\rangle$ and $+$ and $-$ 
denote upper and lower atomic states respectively.

We have numerically solved the master equation for the steady state for 
arbitrary driving field by truncating the Fock basis at a point where the 
population of $|n_{max},\pm\rangle$ is less than $10^{-4}$.  The second order 
correlation function
\begin{equation}
g^{(2)}(\tau) = \frac{\langle a^\dagger(0)a^\dagger(\tau)a(\tau)a(0)\rangle}
{\langle a^\dagger a\rangle^2_{ss}}
\end{equation}
is calculated from steady state matrix elements using the quantum regression 
theorem due to Lax \cite{Lax}.

\subsection{Quantum trajectory simulation}
We have developed a quantum trajectory simulation of this system from the 
master equation following the formalism of Carmichael \cite{Carmichael}.  We 
unravel the master equation into both a piece describing continuous evolution 
and a set of collapse operators in a way which is based on a simulated photon 
counting experiment.
\begin{equation}
{\cal L}\rho = \left ({\cal L - S}\right ) \rho + {\cal S} \rho
\end{equation}
where $\left( {\cal L - S}\right ) \rho$ is identified as the terms which can 
be written as commutators or anticommutators and $\cal S \rho$ is identified 
as all terms which can be written as $\hat O^\dagger\rho\hat O$.  This 
particular unraveling is well suited for studies of photon statistics as the 
$\hat O$'s represent quantum jumps due to emission of a photon.  The 
continuous evolution of the system is described by 
$\left ( {\cal L - S}\right ) \rho$ while $\cal S\rho$ describes collapse 
events which punctuate the evolution.  We define a closed system Hamiltonian 
and a dissipative Hamiltonian from the unraveled master equation as 
\begin{eqnarray}
\left ({\cal L - S}\right ) \rho &=& -\frac{i}{\hbar}\left [ H_S,\rho \right ]
+ \left [ H_D,\rho\right ]_+ \\
&=& -\frac{i}{\hbar}\left [ H_A + H_F + H_{AF} + 
H_L,\rho\right ] - \left [ \left ( a^\dagger a + \sum_j \sigma_+^j\sigma_-^j
\right ),\rho\right ]_+ \nonumber
\end{eqnarray}
where $\left [ \hat O,\rho\right ]_+$ denotes the anticommutator of $\hat O$ 
and $\rho$.  A non-Hermitian Hamiltonian which reproduces the continuous 
evolution of the density matrix is defined as
\begin{equation}
H = H_S + i\hbar H_D.
\end{equation}
The rest of the master equation enters as collapse operators which are applied 
at random times when $R(0,1) < P_c$ where $R(0,1)$ is a random number between 
zero and one and 
\begin{equation}
P_c = \langle \psi|\hat O^\dagger \hat O|\psi\rangle\, dt.
\end{equation}
The time step size is $(20r)^{-1}$ where $r$ is the fastest rate in the 
problem.  In the event that we get two collapse processes in a single time 
step, we use a random number to choose one of the collapses.  The time step 
is chosen so as to minimize such occurrences.
For this system, the collapse operators are $\hat a$ and $\sigma_-^j$ 
corresponding to emission of a photon from the cavity field and the atom 
respectively.  In Section \ref{sec:dephasing}, we describe dephasing due to an 
atom leaving the cavity using another collapse operator.

Because this unraveling of the master equation is based on photon counting 
experiments, the calculation of the second order correlation function is 
carried out quite naturally.  The collapse operator $\hat a$ corresponds to 
emission and detection of a photon from the cavity field mode.  We calculate 
$g^{(2)}(\tau)$ by building up a histogram of delay times between photon 
detections averaged over a long evolution time in a way analogous to 
experimental measurement.

\section{Non-weak driving field}
\label{sec:strong}
The photon statistics of the transmitted light have already been calculated 
in the weak field limit using a truncated five state basis where the system 
has up to two quanta of energy in it \cite{IEEE}.  The three types of 
nonclassical behavior previously discussed have been seen in subsequent 
experiments, however these experiments are not strictly in the weak field 
limit.  It is of interest then to calculate the photon statistics for 
arbitrary driving field and to see to what extent the nonclassical effects 
persist.  It is expected that for a strong enough driving field, the atoms 
saturate and the nonclassical photon correlations will be washed out 
because the cavity will basically contain a coherent state which is only mildly 
perturbed by the presence of the atom.

Looking at the photon correlations in the weak field limit from the point of 
view of quantum trajectories, we can interpret the nonclassical effects as 
resulting from the collapse of the wavefunction.  The detection of the first 
photon emitted from the steady state collapses the wavefunction of the system 
($|\psi_{ss}\rangle \to a|\psi_{ss}\rangle$) and the subsequent time evolution 
as the system returns to the steady state determines the photon correlations.  
The second order correlation function is given by the probability of detecting 
a second photon normalized to the probability of detecting a photon in the 
steady state.  Here we start the system in the steady state, collapse the 
wavefunction, and let it evolve to get \cite{fatpaper}
\begin{equation}
g^{(2)}(\tau) = \frac{\langle a^\dagger(\tau) a(\tau)\rangle_c}{\langle 
a^\dagger a\rangle_{ss}}.
\end{equation}
This assumes that there is usually only one photon emitted as the system 
returns to steady state which is a good approximation in the weak field limit.  
An example of this is shown in Fig. \ref{fig:collapse} for the case of the 
overshoot violation.  Outside of the weak field limit the photon correlation
s are altered for two reasons.  Most simply, the time evolution following a 
collapse from steady state will be altered by the stronger driving field.  
Another effect, however, is the presence of multiple collapses before the 
system returns to the steady state.  Consider a multiple collapse process.  
The first photon comes from the steady state and collapses the wavefunction of 
the system ($|\psi_{collapse1}\rangle = a|\psi_{ss}\rangle$).  Now the time 
evolution occurs as before.  However, the second photon collapses the system 
to a new state which depends on the delay time since the emission of the first 
photon [$|\psi_{collapse2}\rangle = a|\psi_{collapse1}(\tau)\rangle$].  If a 
third photon is emitted before the system returns to steady state then its 
delay time will depend on the details of the evolution from 
$|\psi_{collapse2}\rangle$.  When averaged over many instances, this process 
will wash out the nonclassical effects because of the different evolution 
following different possible states, $|\psi_{collapse2}\rangle$.  An example of 
this process is shown in Fig. \ref{fig:2photon} where the conditioned photon 
number undergoes two collapse events.  This figure shows three delay times of 
$1/\gamma$, $2/\gamma$, and $3/\gamma$ with different evolutions of the 
conditioned photon number resulting in each case.

Figure \ref{fig:evolve} shows the time evolution of $\langle a^\dagger 
a\rangle_c$ following a photon emission from the steady state for a variety of 
system parameters.  The overshoot persists in the evolution of the field 
following emission of a photon from the cavity for a driving field as large as 
$E/E_{sat} = 0.41$.  The undershoot and sub-Poissonian statistics survive for 
driving fields as large as $E/E_{sat} = 0.8$ and $E/E_{sat} = 0.37$ 
respectively.  The saturation field strength $E_{sat}$ is the driving field 
for which 
\begin{equation}
\langle n\rangle = n_{sat} = \frac{\gamma^2}{8g^2}.
\end{equation}

The photon statistics of the transmitted field are shown in Fig. 
\ref{fig:strong} for the three types of nonclassical effects seen in this 
system at a variety of driving field intensities.  Figure \ref{fig:strong}(a) 
shows $g^{(2)}(\tau)$ for system parameters ($g/\gamma = 1, \kappa/\gamma = 
0.77$) which produce an overshoot violation of the Schwarz inequality 
[$g^{(2)}(\tau) > g^{(2)}(0)$] in the weak field limit.  At a driving field of 
$E/E_{sat} = 0.17$ the overshoot violation is gone, thus this nonclassical 
effect is quite dependent on the weak driving field.  Figure 
\ref{fig:strong}(b) shows photon statistics for system parameters ($g = 
2/\gamma, \kappa/\gamma = 5$) which produce an undershoot violation of the 
Schwarz inequality [$1 - g^{(2)}(\tau)_{min} > g^{(2)}(0) - 1$] in the weak 
field limit.  Here the nonclassical effect disappears at a driving field of 
$E/E_{sat} = 0.28$ showing that this is a more robust effect.  Figure 
\ref{fig:strong}(c) shows photon statistics for system parameters ($g/\gamma 
= 1, \kappa/\gamma = 1.6$) which produce photon antibunching [$g^{(2)}(0)_+ > 
g^{(2)}(0)$] and sub-Poissonian statistics [$g^{(2)}(0) < 1$] in the weak 
field limit.  In this case the nonclassical effect persists until $E/E_{sat} 
= 0.16$ where the system shows slight bunching and super-Poissonian 
statistics.  (Notice that the nonclassical effects are not as robust as the 
time evolution of the cavity field would indicate.  Therefore the destruction 
of nonclassical effects are in part a result of multiple photon processes.)  
For all system parameters the transmitted light becomes super-Poissonian as 
the driving field is increased.

\section{Atomic transit dephasing}
\label{sec:dephasing}
We now turn our attention to the effects of atomic traversal of the cavity on 
the photon statistics.  In previous work it has been assumed that the atoms 
are all fixed at antinodes of the cavity field and so had the maximum coupling 
$g_0 = \mu \left ( \omega_c/2\hbar\epsilon_0 V \right )^{1/2}$.  Experiments 
on this system have used atomic beams to send atoms through a cavity.  This 
atomic traversal of the cavity will introduce two new effects.  First, the 
atom/cavity field coupling will depend on the position of the atom in the 
cavity.  One might think that this would destroy the nonclassical 
correlations.  However, the atoms with the largest coupling interact most 
strongly with the field and are most likely to contribute to the correlations.  
So the atoms near an antinode will have the largest contribution and other 
atoms may have little effect on the correlations.  This issue will be further 
addressed in Sec. \ref{sec:two atoms}.  The second effect of atomic traversal 
is dephasing which occurs when an atom enters or leaves the cavity.  It is 
this effect which we consider in this section.

We have used two approaches to model the dephasing due to atomic traversal.  
The first is to add a term to the master equation which describes nonradiative 
decay of atomic polarization.
\begin{equation}
\dot\rho = {\cal L}\rho + \gamma_{ph}\left (\sigma_z\rho\sigma_z - \rho\right )
\end{equation}
This term in the master equation has its origins in collisional processes 
\cite{Howardbig} and so may or may not accurately describe the dephasing which 
occurs when an atom leaves the cavity.

The second approach uses a quantum trajectory simulation of the system to 
model the dephasing.  In this approach we assume that there is always exactly 
one atom in the cavity.  An atom leaves the cavity and another atom enters the 
cavity in the ground state at a rate $\gamma_{ph}$.  We can assume the atom 
enters the cavity in the ground state, but it is not immediately clear how to deal 
with the state of the exiting atom.  This atom is in some superposition of 
excited and ground state and these states are entangled with the cavity field 
state.  One approach would be to leave the photon number distribution of the 
cavity field unchanged using a collapse operator which has the following 
action on the state of the system:
\begin{equation}
|\psi\rangle = \sum_n\left ( c_{e,n}|e,n\rangle + c_{g,n}|g,n\rangle\right )
\to |\psi_c\rangle = \sum_n\left ( c_{e,n}^2 + c_{g,n}^2\right ) |g,n\rangle.
\end{equation}
However, this is not a consistent application of the quantum trajectories.  
Consider the evolution of the atom after it leaves the cavity.  The atom at 
some later time may emit a photon into the vacuum meaning it was in the 
excited state when it left the cavity; or it will never emit a photon, 
meaning it was in the ground state when it left the cavity.  In general, the 
atom and environment and by entanglement the atom/cavity system will then be 
described by a density operator.  However, we wish to use a pure state to 
describe the atom/cavity system conditioned on the detection of transmitted 
photons.  To be consistent we must use a pure state to describe the atom after 
it has left the cavity.  We use a collapse operator which picks either the 
excited state field distribution or the ground state field distribution of the 
system and then places the new atom in the ground state.  This operator has 
the following action:
\begin{mathletters}
\begin{eqnarray}
|\psi\rangle = \sum_n\left ( c_{e,n}|e,n\rangle + c_{g,n}|g,n\rangle\right )
\to |\psi_c\rangle = \sum_n c_{e,n}|g,n\rangle \qquad {\rm with\ probability}\  
c_{e,n}^2 \\
|\psi\rangle = \sum_n\left ( c_{e,n}|e,n\rangle + c_{g,n}|g,n\rangle\right )
\to |\psi_c\rangle = \sum_n c_{g,n}|g,n\rangle \qquad {\rm with\ probability}\  
c_{g,n}^2. 
\end{eqnarray}
\end{mathletters}
This collapse operator is then applied at a Gaussian distributed series of 
times with average $1/\gamma_{ph}$ and a full width of $1/\gamma_{ph}$ as this 
approximates the traversal times of atoms with a Maxwell-Boltzmann velocity 
distribution.  This model of dephasing differs from collisional dephasing in 
two important ways.  First, it does not enter the deterministic Hamiltonian 
evolution between collapses at all, whereas the collisional dephasing in the 
trajectory picture would have both a collapse component as well as the 
decay of coherence in the continuous evolution.  Second, this dephasing always 
places the atom into the ground state whereas the collisional dephasing places 
the atom in the ground or excited state with probabilities determined by the 
relative populations.

Now we turn to our results and a comparison of the two types of dephasing.  
Because the traversal dephasing does not affect the deterministic evolution of 
the system we expect that for a given dephasing rate it will be less 
destructive of the nonclassical photon statistics than collisional dephasing.  
In Fig. \ref{fig:dephasing} we show the second order correlation function 
with collisional dephasing.  All three types of nonclassical effects are quite 
sensitive to this dephasing with $\gamma_{ph} = 0.05$ destroying the overshoot 
[Fig. \ref{fig:dephasing}(a)] and undershoot [Fig. \ref{fig:dephasing}(b)] 
while the sub-Poissonian statistics survive until $\gamma_{ph} = 0.2$ in Fig. 
\ref{fig:dephasing}(c).  Our results for the transit dephasing are shown in 
Fig. \ref{fig:dephasing2}.  The overshoot in Fig. \ref{fig:dephasing2}(a) is 
again extremely sensitive to dephasing with classical statistics for 
$\gamma_{ph}/\gamma = 0.05$ while the sub-Poissonian statistics in Fig. 
\ref{fig:dephasing2}(b) are more robust, surviving up to $\gamma_{ph}/\gamma 
= 0.5$.

\section{Two atom effects}
\label{sec:two atoms}
We now consider the effect of placing two atoms inside the cavity either both 
at antinodes of the field, or allowing one of the atoms to be arbitrarily 
placed so that its coupling is in the range $0$ to $g_0$.  Dephasing is not 
considered in this Section.  In the experiments conducted on this system it is 
likely that there is some effect from 'spectator atoms' which are located away 
from an antinode of the field and so do not contribute to the nonclassical 
photon statistics.  If there are enough of these atoms they may simply absorb 
light and emit it out of the cavity thus effectively decreasing the quality of 
the cavity.  As a first step towards understanding the effect of spectator 
atoms, we place one extra atom in the cavity with a coupling which is some 
fraction of the original atom's coupling.  We use a quantum trajectory 
simulation to calculate photon statistics.

In Fig. \ref{fig:spectator} we plot $g^{(2)}(\tau)$ while allowing the 
coupling of the spectator atom ($g_2$) to vary from $g_0/10$ to $g_0$.  We see 
that for all three sets of parameters the photon statistics vary continuously 
from the single atom result to the two atom result with no qualitative 
deviation in the photon statistics.  This suggests that the spectator atoms 
can have an observable effect on the statistics, but they do not destroy the 
nonclassical correlations at least when close to the ideal condition of having 
a single atom in the cavity at a time.  A group of many spectator atoms may be 
more detrimental to nonclassical correlations.

We now consider the case in which both atoms are placed at antinodes of the 
field.  The photon statistics for this system have been solved for an arbitrary 
number of maximally coupled atoms using a set of symmetrized Dicke states to 
describe the atomic excitation \cite{fatpaper}.  This would correspond to an 
experimental setup where it is not possible to tell which atom spontaneously 
emitted a given photon.  However, there would be experimental situations where 
the symmetrized states are not valid states with respect to spontaneous 
emission.  Here, using quantum trajectories, we consider both a symmetrized 
collapse and an unsymmetrized collapse for spontaneous emission.  The operator 
for the Dicke collapse is
\begin{equation}
\hat C_{Dicke} = \frac{1}{\sqrt{2}}\left ( \sigma_-^1 + \sigma_-^2 \right )
\end{equation}
so that when a spontaneous emission event occurs, both atoms are collapsed 
symmetrically.  For the non-Dicke collapse the atomic operators are used 
separately so that one atom or the other collapses.

For weak driving field we expect that there will be no difference between 
these types of collapse because in this limit we only detect photons emitted 
from steady state.  However, for stronger driving fields we begin to detect 
photons emitted from the collapsed state and the two collapses give a 
different collapsed state ($\sigma_-^1|\psi_{ss}\rangle$ or 
$\sigma_-^2|\psi_{ss}\rangle$ versus $\hat C_{Dicke}|\psi_{ss}\rangle$), and 
therefore, different photon statistics.

In Fig. \ref{fig:dicke} $g^{(2)}(\tau)$ is plotted for the two types of 
collapse for a driving field of $E = 0.5$.  Figure \ref{fig:dicke}(a) shows a 
significant difference as the nonclassical statistics are completely gone for 
the non-symmetrized collapse.  Figure \ref{fig:dicke}(b) shows no dependence 
on the type of collapse.  Figure \ref{fig:dicke}(c) shows a mild dependence 
on the type of collapse with a slightly larger value of $g^{(2)}(0)$ for the 
non-symmetrized collapse.

\section{Conclusion}
\label{sec:conclude}
We have investigated extensions to previous theoretical work on a driven 
atoms/cavity system with dissipation.  We have calculated the normalized 
second order correlation function for the transmitted light including effects 
of arbitrary driving fields, non-radiative dephasing, and arbitrary coupling 
strength for multiple atoms.  We have found that nonclassical field states are 
easily destroyed by deviations from the weak field limit and by non-radiative 
dephasing modeled as both collisional dephasing and as atomic transit 
dephasing.  We have also found that allowing two atoms in the cavity with 
different atom/field coupling strengths does not have a detrimental effect on 
the nonclassical field.  The experiments which have been done on this system 
have not really in the weak field limit.  As $E \to 0$ the number of counts 
also goes to zero so it is difficult to carry out experiments in this regime, 
but this work suggests that it is important.

\section*{Acknowledgments}
The authors would like to thank Howard J. Carmichael, Luis Orozco, and Greg 
Foster for many useful conversations.



%
%
\begin{figure}
\caption{Examples of nonclassical photon statistics in the weak field limit.  
(a) shows sub-Poissonian statistics and photon antibunching.  (b) shows an 
overshoot violation of the Schwarz inequality.  (c) shows an undershoot 
violation of the Schwarz inequality.}
\label{fig:weak}
\end{figure}

\begin{figure}
\caption{A diagram of the system.  The cavity field decays at a rate 
$\kappa$, the atom spontaneously emits at a rate $\gamma$, and there is an 
electric dipole coupling between the atom and cavity field with a strength 
$g$.}
\label{fig:diagram}
\end{figure}

\begin{figure}
\caption{The conditioned photon number normalized by the steady state photon 
number for the cavity field.  This is identical to $g^{(2)}(\tau)$ in the weak 
field limit.  Parameters are $g/\gamma = 1, \kappa/\gamma = 0.77, E/\gamma = 
0.01$.}
\label{fig:collapse}
\end{figure}

\begin{figure}
\caption{The conditioned photon number normalized by the steady state photon 
number for the cavity field.  In these plots, two collapses occur giving rise 
to different evolutions depending on the delay between collapses.  The delays 
between collapses are (a) $1/\gamma$, (b) $2/\gamma$, and (c) $3/\gamma$.  On 
the average, processes such as these will wipe out the nonclassical 
correlations.}
\label{fig:2photon}
\end{figure}

\begin{figure}
\caption{Time evolution of the conditioned cavity photon number following 
emission of a photon from the cavity and collapse of the wavefunction from 
steady state.  The plots are for (a) $g/\gamma = 1, \kappa/\gamma = 0.77, 
E/\gamma$ = 0.1 (solid line), 0.2 (dashed line), 0.3 (dotted line).  (b) 
$g/\gamma = 2, \kappa/\gamma = 5, E/\gamma$ = 0.1 (solid line), 0.5 (dashed 
line), 1 (dotted line).  (c) $g/\gamma = 1, \kappa/\gamma = 1.6, E/\gamma$ = 
0.1 (solid line), 0.5 (dashed line), 1 (dotted line).}
\label{fig:evolve}
\end{figure}

\begin{figure}
\caption{Photon statistics of the transmitted field for varying driving field 
strength.  The plots are for (a) $g/\gamma = 1, \kappa/\gamma = 0.77, 
E/\gamma$ = 0.025 (solid line), 0.125 (dashed line), 0.2 (dotted line), 0.35 
(dash-dotted line).  (b) $g/\gamma = 2, \kappa/\gamma = 5, E/\gamma$ = 0.025 
(solid line), 0.25 (dashed line), 0.35 (dotted line), 0.5 (dash-dotted line), 
1 (small-dashed line).  (c) $g/\gamma = 1, \kappa/\gamma = 1.6, E/\gamma$ = 
0.025 (solid line), 0.25 (dashed line), 0.425 (dotted line), 0.6 (dash-dotted 
line).}
\label{fig:strong}
\end{figure}

\begin{figure}
\caption{Photon statistics of the transmitted field with collisional 
dephasing.  The plots are for (a) $g/\gamma = 1, \kappa/\gamma = 0.77, 
E/\gamma = 0.1, \gamma_{ph}/\gamma$ = 0 (solid line), 0.05 (dashed line), 0.2 
(dotted line).  (b) $g/\gamma = 2, \kappa/\gamma = 5, E/\gamma = 0.1, 
\gamma_{ph}/\gamma$ = 0 (solid line), 0.05 (dashed line), 0.2 (dotted line).  
(c) $g/\gamma = 1, \kappa/\gamma = 1.6, E/\gamma = 0.1, \gamma_{ph}/\gamma$ = 
0 (solid line), 0.1 (dashed line), 0.2 (dotted line).}
\label{fig:dephasing}
\end{figure}

\begin{figure}
\caption{Photon statistics of the transmitted field with atomic transit 
dephasing.  The plots are for (a) $g/\gamma = 1, \kappa/\gamma = 0.77, 
E/\gamma = 0.1, \gamma_{ph}/\gamma$ = 0.05 (solid line), 0.1 (dashed line), 
0.5 (dotted line).  (b) $g/\gamma = 1, \kappa/\gamma = 1.6, E/\gamma = 0.1, 
\gamma_{ph}/\gamma$ = 0.05 (solid line), 0.1 (dashed line), 0.5 
(dotted line).}
\label{fig:dephasing2}
\end{figure}

\begin{figure}
\caption{Photon statistics of the transmitted field with two atoms in the 
cavity at arbitrary coupling strength.  The plots are for (a) $g_1/\gamma = 
1, \kappa/\gamma = 0.77, E/\gamma = 0.1, g_2/\gamma$ = 0.1 (solid line), 0.5 
(dashed line), 1 (dotted line).  (b) $g_1/\gamma = 2, \kappa/\gamma = 0.77, 
E/\gamma = 0.1, g_2/\gamma$ = 0.2 (solid line), 1 (dashed line), 2 (dotted 
line).  (c) $g_1/\gamma = 1, \kappa/\gamma = 1.6, E/\gamma = 0.1, g_2/\gamma$ 
= 0.1 (solid line), 0.5 (dashed line), 1 (dotted line).}
\label{fig:spectator}
\end{figure}

\begin{figure}
\caption{Photon statistics for the transmitted field with two atoms comparing 
the symmetrized Dicke spontaneous emission with non-symmetrized spontaneous 
emission.  The plots are for (a) $g_1/\gamma = g_2/\gamma = 1, \kappa/\gamma 
= 0.77, E/\gamma$ = 0.1.  (b) $g_1/\gamma = g_2/\gamma = 2, \kappa/\gamma = 
5, E/\gamma$ = 0.1.  (c) $g_1/\gamma = g_2/\gamma = 1, \kappa/\gamma = 1.6, 
E/\gamma$ = 0.1.  The solid line is the non-symmetrized collapse and the 
dashed line is the symmetrized collapse.}
\label{fig:dicke}
\end{figure}

%
%

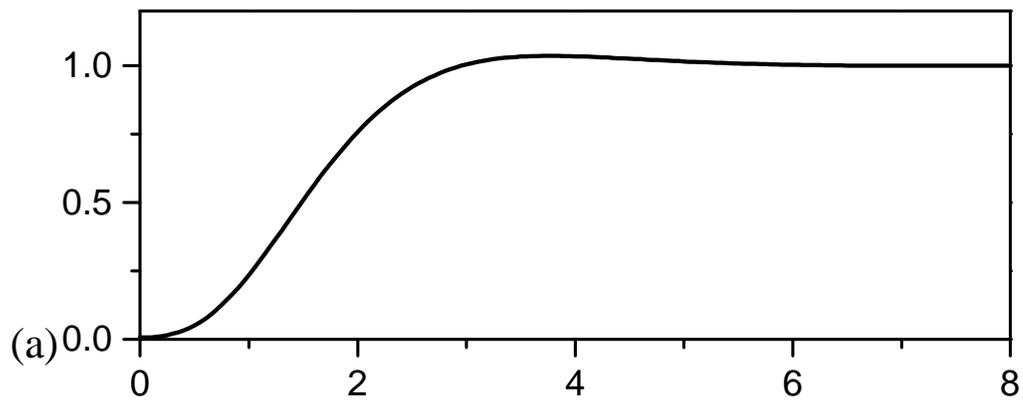
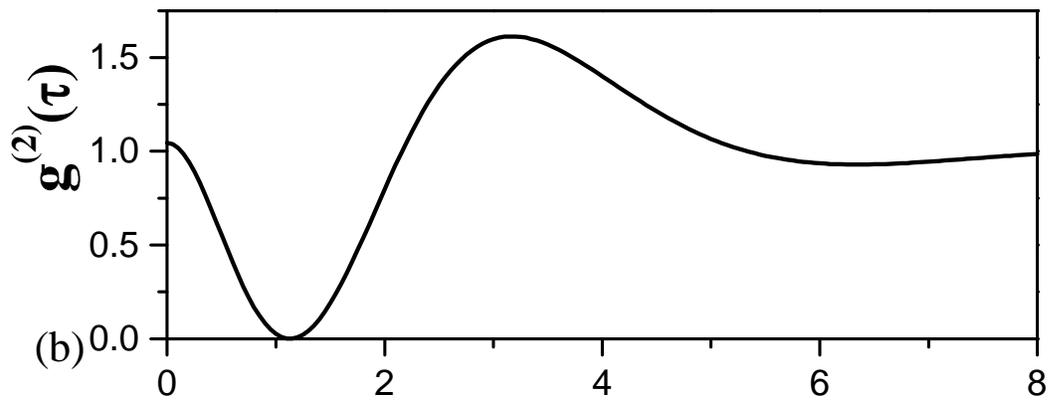
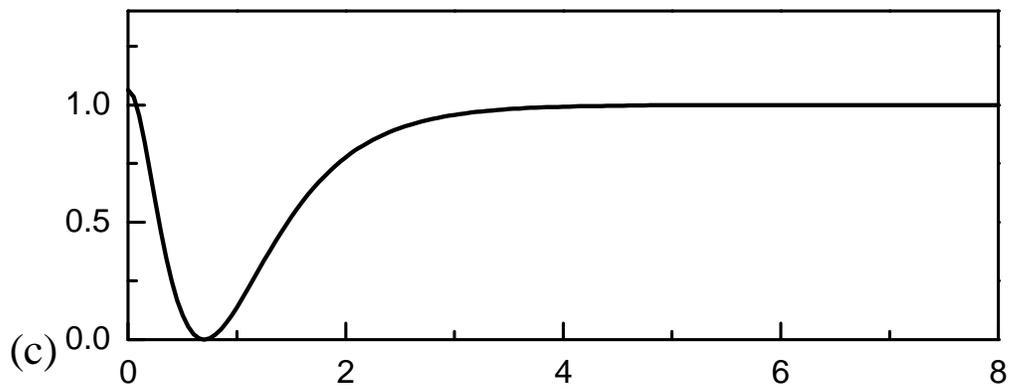
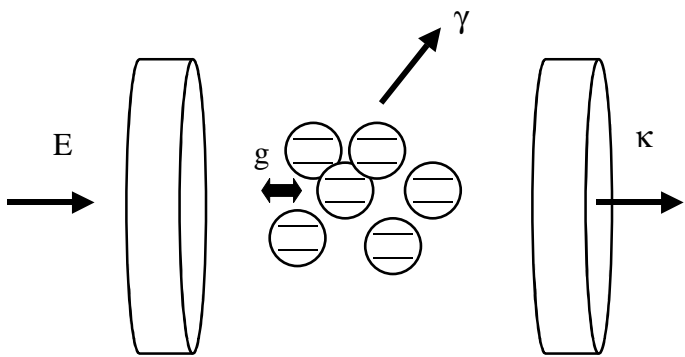
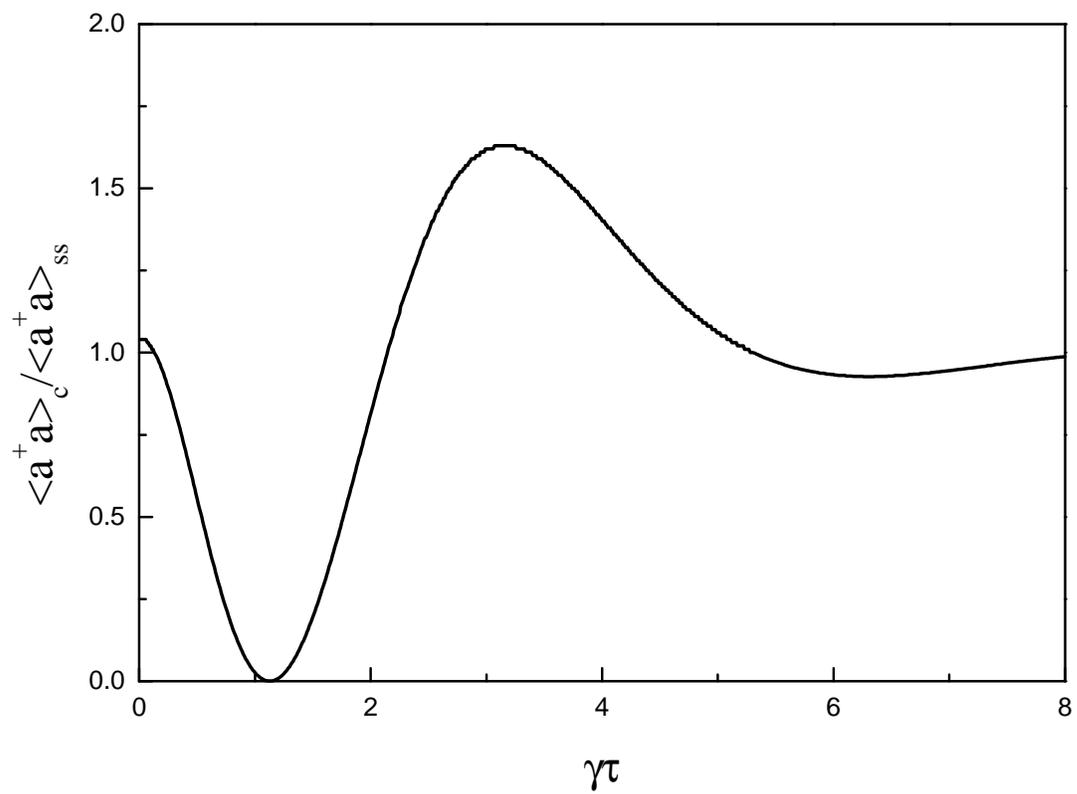
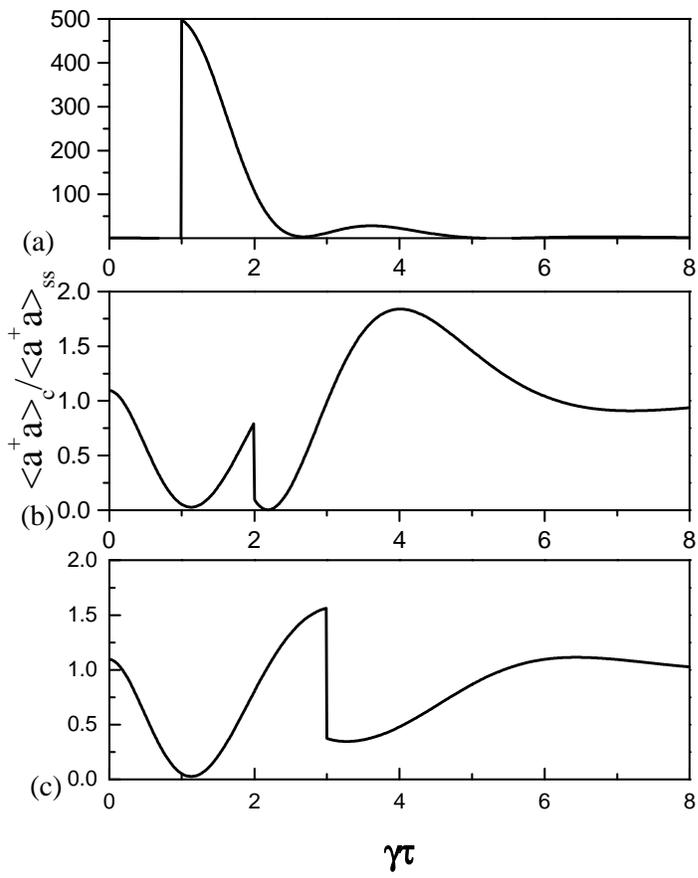
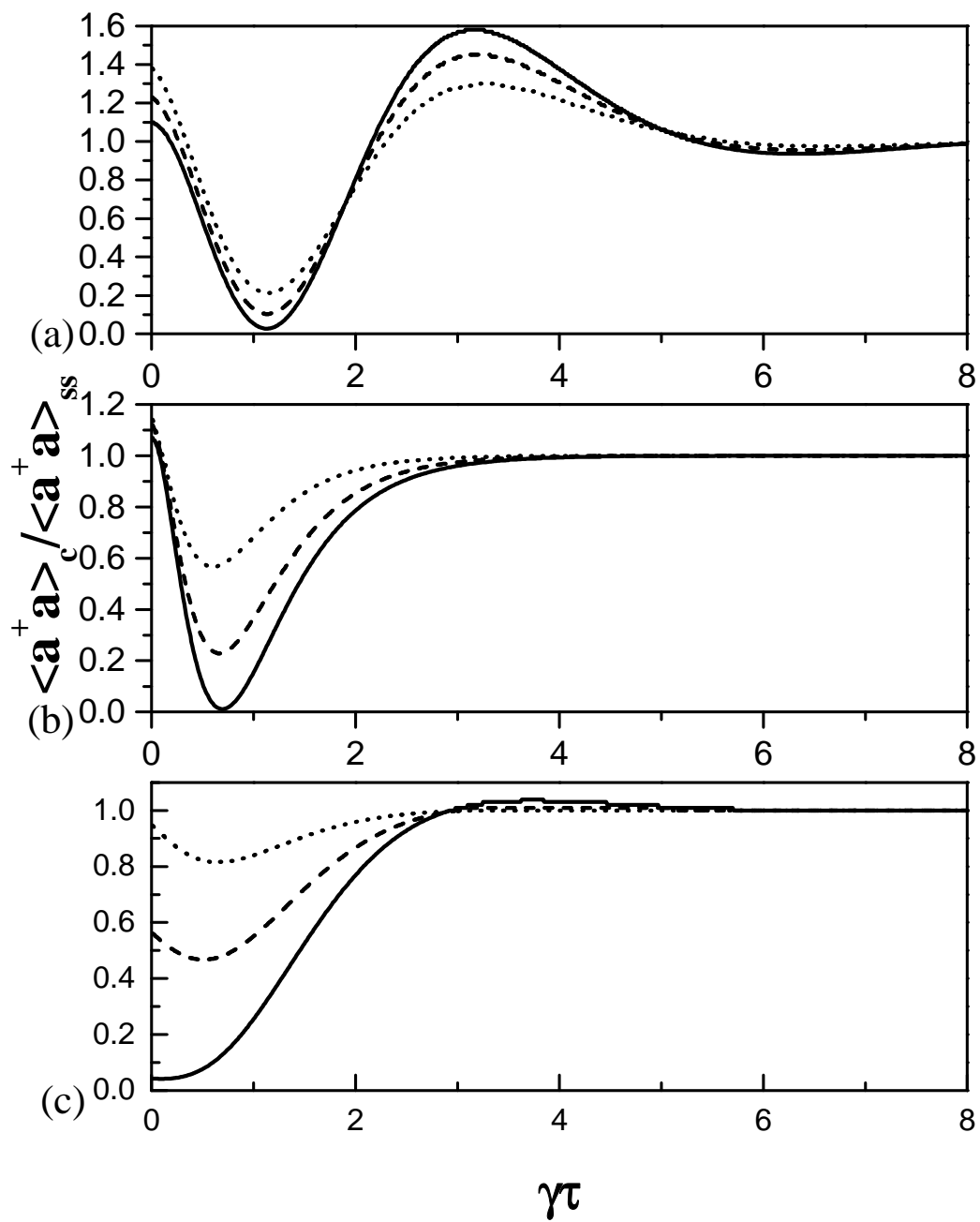
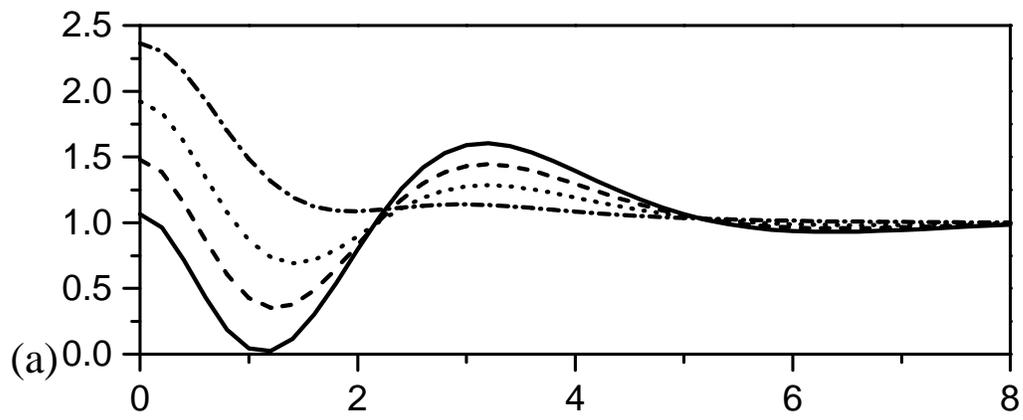
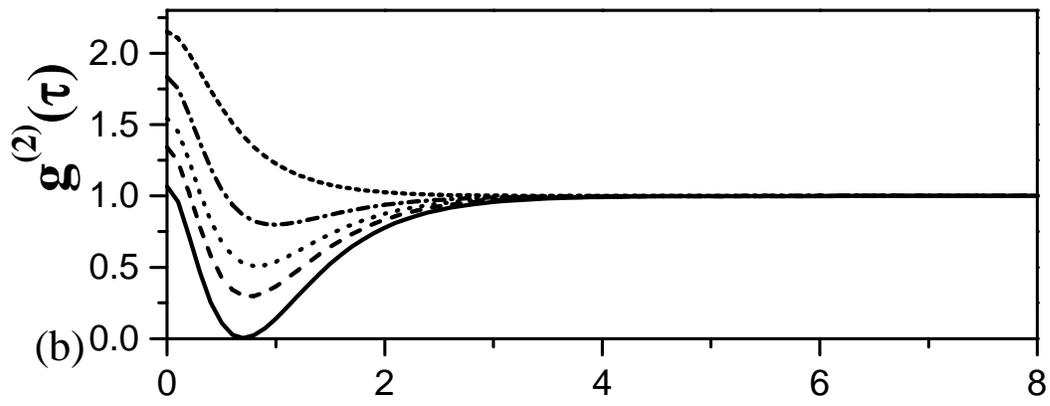
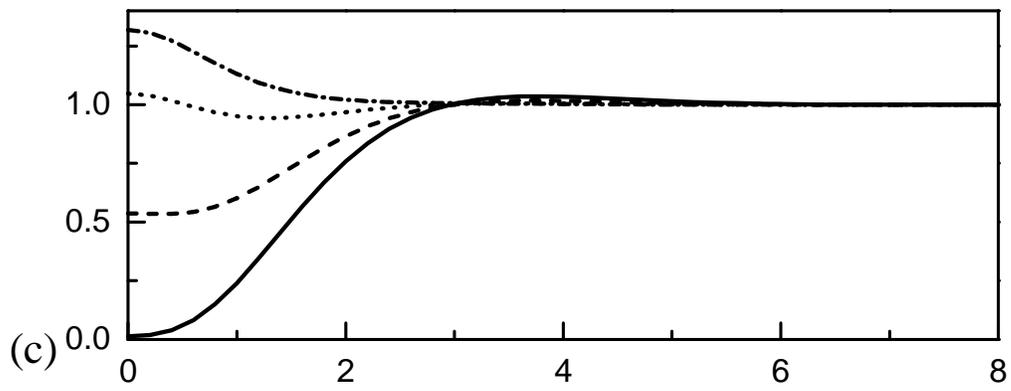
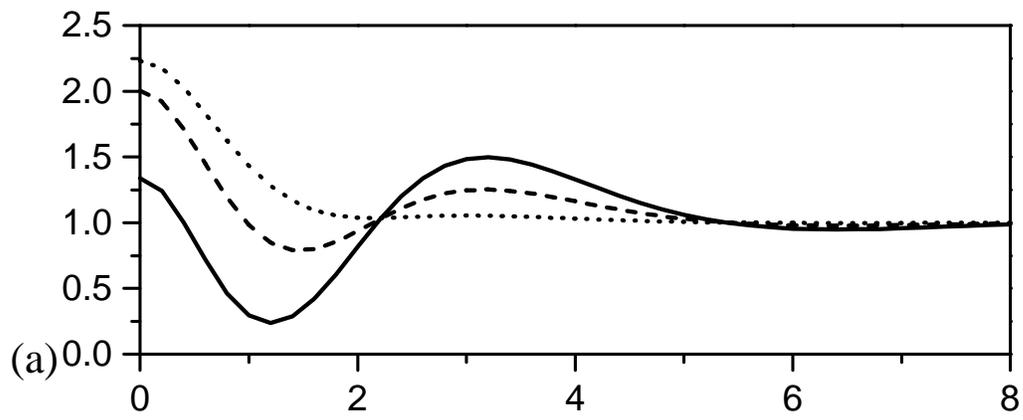
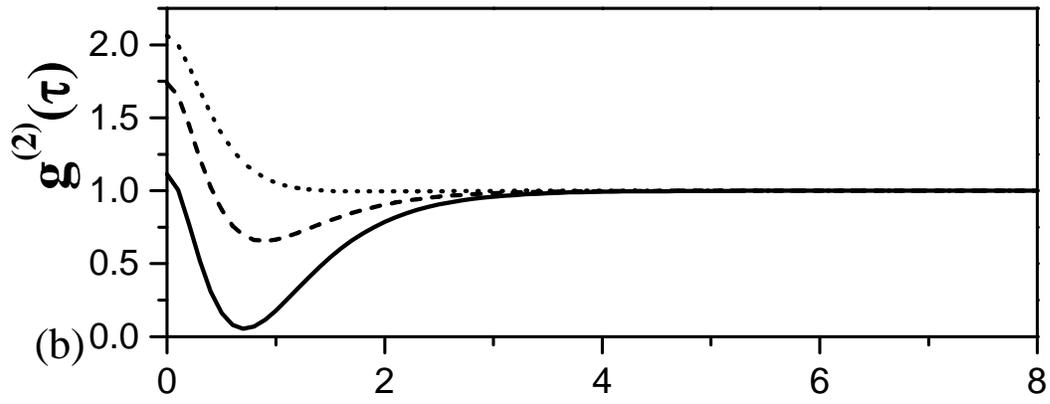
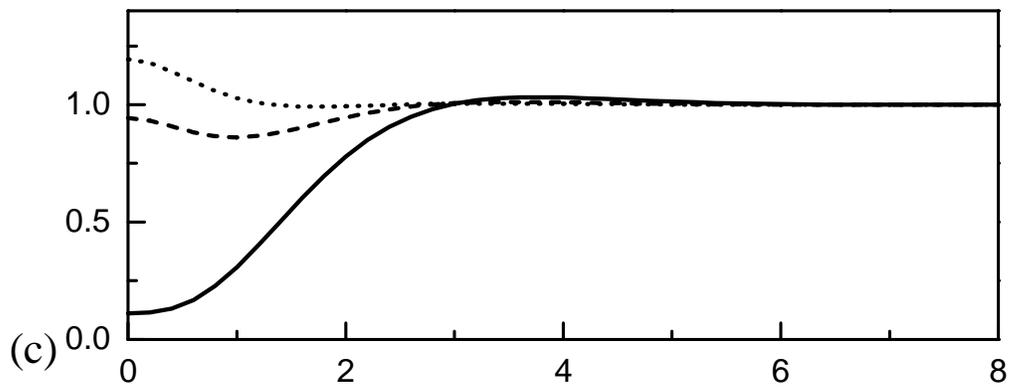
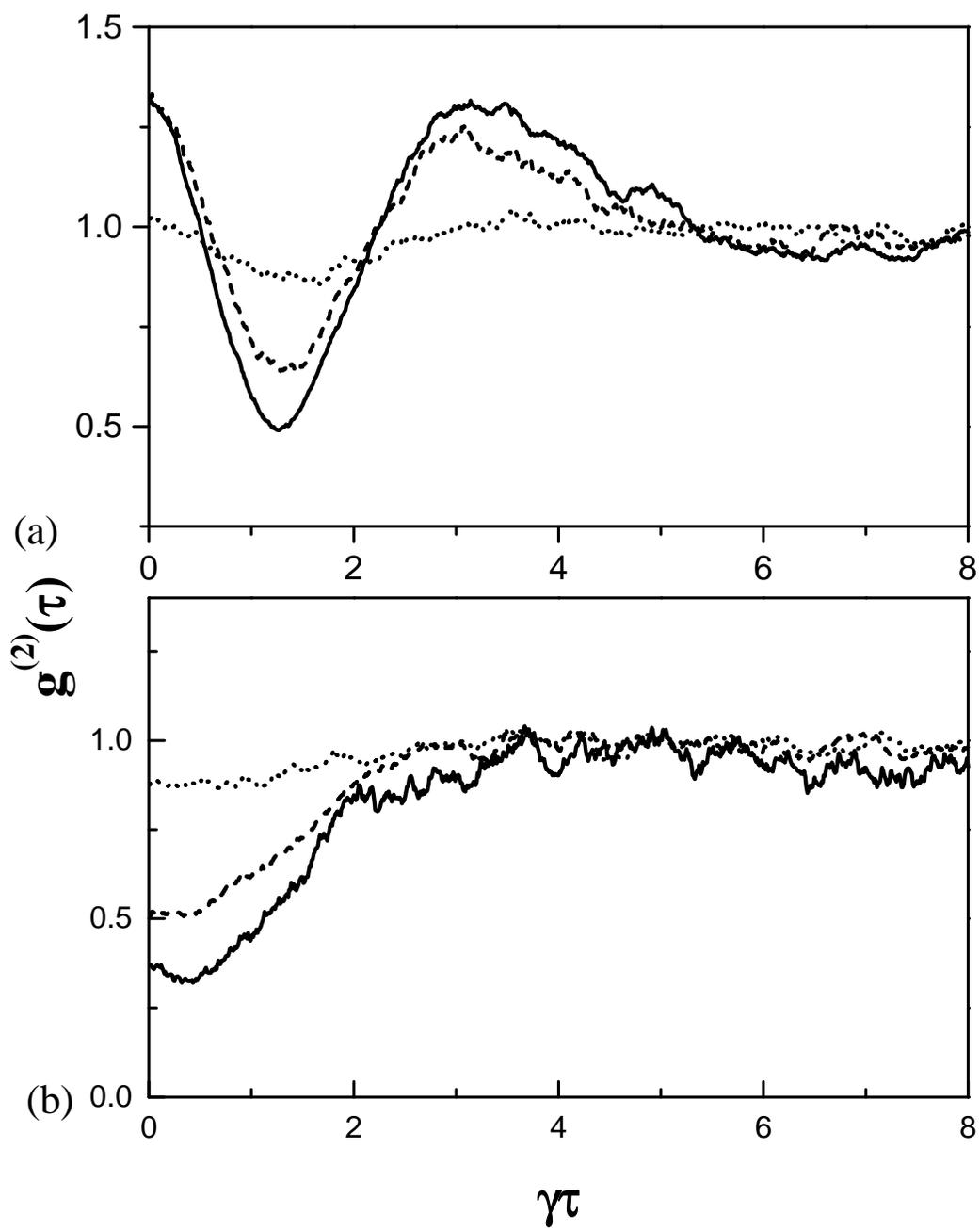
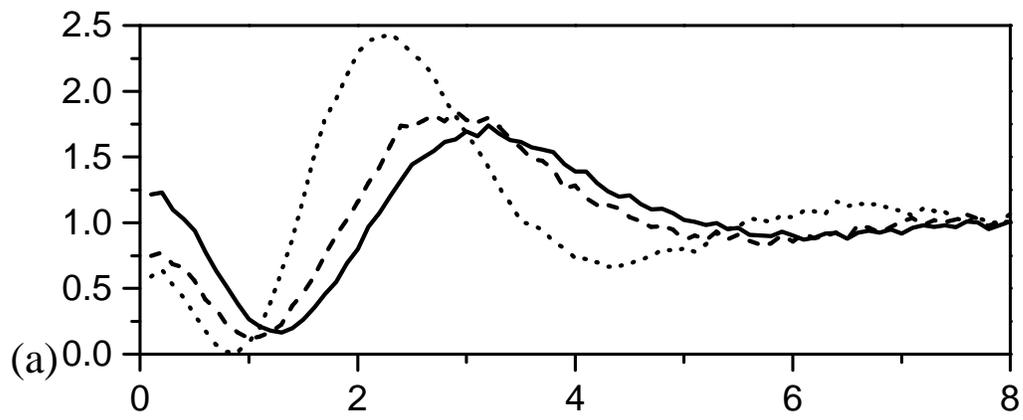
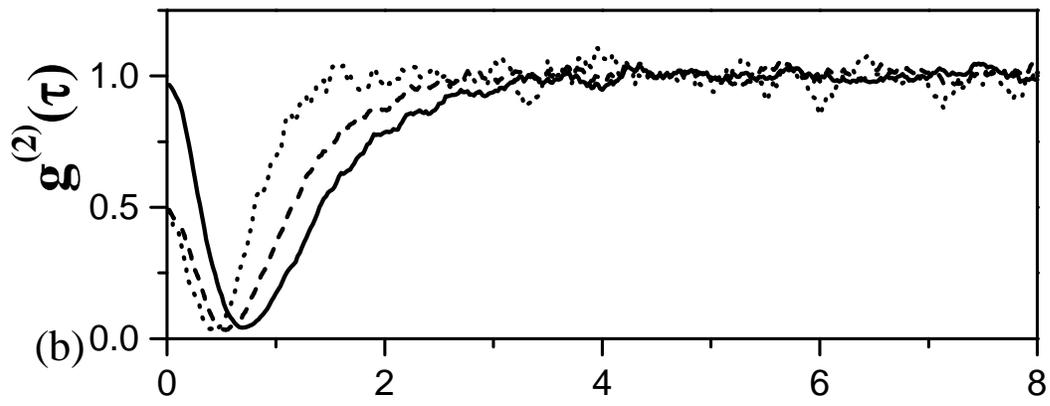
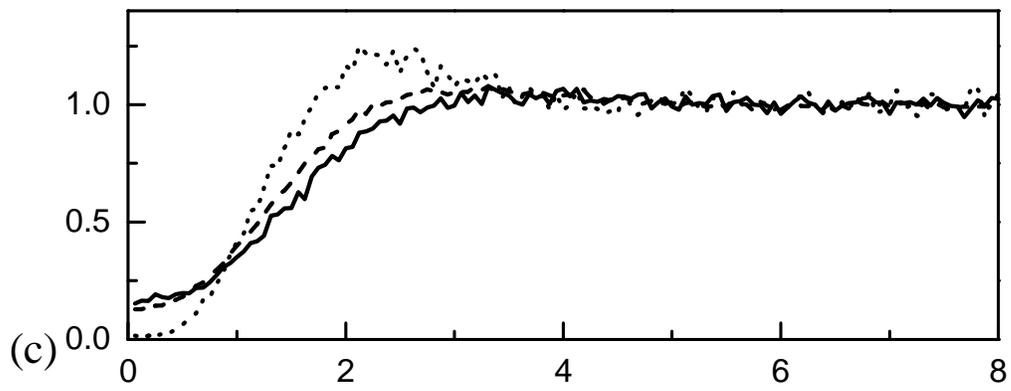
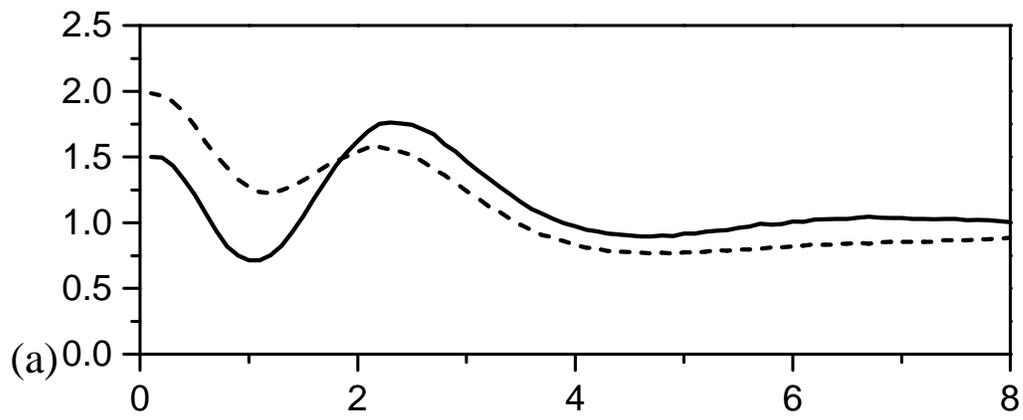
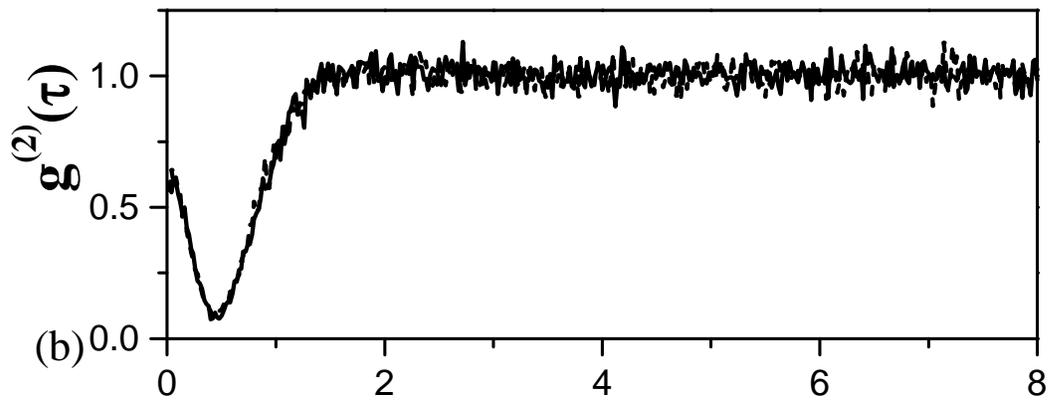
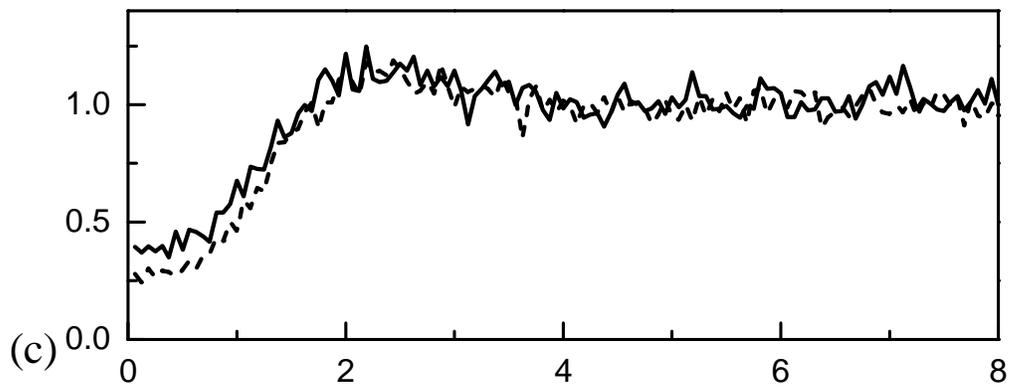
\end{document}